\documentclass[10pt,a4paper]{article} 
\usepackage[dvips]{graphicx}
\usepackage{amsfonts,amsmath,amssymb}
\usepackage{hyperref}
\usepackage{cite}
\usepackage{enumitem}
\numberwithin{equation}{section}
\newcommand{\bse}{\begin{subequations}}
\newcommand{\ese}{\end{subequations}}

\begin{document}
\title{\bf Confining D-Instanton Background in an External Electric Field}

\date{}
\maketitle
\vspace*{-0.3cm}
\begin{center}
{\bf Leila Shahkarami$^{a,1}$, Farid Charmchi$^{b}$}\\
\vspace*{0.3cm}
{\it {${}^a$School of Physics, Damghan University, Damghan, 41167-36716, Iran}}\\
{\it {${}^b$Research Institute for Astronomy and Astrophysics of Maragha (RIAAM), P.O.Box 55134-441, Maragha, Iran}} \\
\vspace*{0.3cm}
{\it  {${}^1$l.shahkarami@du.ac.ir}}
\end{center}

\begin{abstract}
Using holography, we discuss the effects of an external static electric field on the D3/D-instanton theory at zero-temperature, which is a quasi-confining theory, with confined quarks and deconfined gluons. 
We introduce the quarks to the theory by embedding a probe D7-brane in the gravity side, and turn on an appropriate $U(1)$ gauge field on the flavor brane to describe the electric field.
Studying the embedding of the D7-brane for different values of the electric field, instanton density and quark masses, we thoroughly explore the possible phases of the system.
We find two critical points in our considerations.
We show that beside the usual critical electric field present in deconfined theories, there exists another critical field, with smaller value, below which no quark pairs even the ones with zero mass are produced and thus the electric current is zero in this (insulator) phase. 
At the same point, the chiral symmetry, spontaneously broken due to the gluon condensate, is restored which shows a first order phase transition.
Finally, we obtain the full decay rate calculating the imaginary part of the DBI action of the probe brane and find that it becomes nonzero only when the critical value of the electric field is reached. 
\end{abstract}
\section{Introduction}
The use of the AdS/CFT correspondence and in general gauge/gravity duality \cite{Maldacena1,Maldacena2,Maldacena3,Solana} have led to major advances in understanding different aspects of strongly-coupled gauge theories.
An important example is how such systems respond to external electromagnetic fields.
Among other things, this study can be useful to understand the confinement/deconfinement phase transition in the RHIC and LHC experiments, where strong electromagnetic fields are created due to heavy ion collisions \cite{em}.

It is known that an external electric field can make 
the vacuum unstable against the production of particle-antiparticle pairs.
This phenomenon known as the Schwinger effect is ubiquitous in any quantum field theory coupled with a $U(1)$ gauge field and is also generalized to nonAbelian gauge theories such as the quantum chromodynamics (QCD).
The Schwinger effect has been evaluated first under weak-coupling and weak-field conditions \cite{Schwinger} and then generalized to the case of arbitrary coupling \cite{manton1}.
These calculations show that beyond a critical electric field $E_c$ the vacuum decays catastrophically, i.e., the potential barrier for the pairs vanishes and they are produced without any obstruction.
The value obtained for $E_c$ is far beyond the weak-field approximation.
However, the existence of such a critical electric field was verified later by studying the Schwinger effect beyond the weak-field approximation using the AdS/CFT correspondence \cite{semenoff}, motivated by its connection to the string theory which predicts the existence of an upper critical electric field \cite{string1,string2}.
Since then, there has been a growing interest in studying the holographic Schwinger effect in different situations, and in both deconfined and confined phases \cite{potential,Sch1,Sch2,Sch3,confin1,confin2,Sch4,Sch5,Sch6,Sch7,confinrev,dehghani}.
In these problems, they consider a probe D3-brane in an intermediate position in the bulk to describe the quarks of finite mass.

The study of the Schwinger effect in confining backgrounds reveals that there exists another critical electric field, $E_s$, below which no pairs even the massless ones can be produced.
For deconfining backgrounds $E_s=0$, i.e., the pair production occurs in the presence of any nonzero external electric field.

In \cite{dehghani} we employed a quasiconfining gauge theory to elaborate more on the response of the theories with confinement to external electric fields.
We chose the D3/D-instanton background first suggested in \cite{instanton1}, where they consider the near horizon limit of the D-instanton charge uniformly distributed on the D3-brane at zero temperature.
This geometry is dual to the ${\cal N}=4$ super Yang-Mills theory with constant gluon condensate.
This theory is called ``quasiconfining'', since it has been verified using AdS/CFT \cite{instanton1} that it is partially confining with confined quarks and deconfined gluons.
The quark-antiquark potential in this theory is found to be linearly rising for large distances \cite{linearP}, showing the confinement of the quarks.
At finite temperature, however, this theory becomes deconfined \cite{linearP,instanton2}, i.e., the transition to the deconfined phase occurs at $T=0$, unlike the QCD.
In \cite{dehghani} we investigated this theory at both zero and nonzero temperatures and found the critical electric fields $E_c$ and $E_s$.
As expected, $E_s$ is only nonzero for the zero-temperature case, i.e., the confined case.

The present paper is devoted to considering the effect of the electric 
field on this confining theory using a completely different method.
Our approach here is to introduce the fundamental quarks to the theory by embedding a probe D7-brane in the dual gravity background.
To describe the external electric 
field, we turn on a nontrivial $U(1)$ gauge field in an appropriate direction of the D7-brane world-volume and find the possible solutions of the DBI action. 
Using this setup, we are able to investigate the behavior of our confining system under external electric fields 
and explore its possible phases.
Our calculations show that the quark condensate can be regarded as an order parameter of phase transitions due to the presence of the electric field.
We moreover compare the results with those of \cite{dehghani}.
We also evaluate the rate of the decay caused by the presence of the electric field on the theory at zero temperature.
This can be done by calculating the imaginary part of the effective DBI action of the probe brane, which can be served as the effective Lagrangian of the theory, using the gauge/gravity duality (see for example \cite{decay1,decay2}).
We should mention here that using a D7-brane probing $N$ D3-branes
, one can study the full decay rate due to turning on the electric field. In contrast, in \cite{dehghani} and the previous similar calculations where a D3-brane is probing $N$ D3-branes, the studies are only in the Coulomb branch with the gauge group $U(N+1)$ spontaneously broken to $U(1)\times U(N)$ and contain just the leading exponent corresponding to the on-shell action of the instanton.

In the next section we consider solutions of a D7 probe brane embedded in the D3/D-instanton theory and briefly discuss about the chiral symmetry in this system.
Then, we turn on an external electric field in section \ref{s3} and consider in detail the response of the system to this field.
Section \ref{imagi} is devoted to the calculation of the imaginary part of the effective Lagrangian for the massless case.
We finally summarize and draw some conclusions in section \ref{conc}.
\section{Review on probe D7-brane in D-instanton background}\label{backgr geo}
In this section we present the setup of a D7 probe brane in the background geometry of our interest and quickly review the effect of the gluon condensate on the chiral symmetry of the system.

We are interested in studying the near horizon geometry of the D3/D-instanton theory
 which is a solution of a ten-dimensional supergravity action which in the Einstein frame is given by \cite{instanton2}
\begin{align}\label{action10D}
 S=\frac{1}{\kappa}\int d^{10}x \sqrt{-g} \left(R-\frac{1}{2}(\partial\Phi)^2 +\frac{1}{2}e^{2\Phi}(\partial\chi)^2-\frac{1}{6}F^2_{(5)} \right).
\end{align}
Here $\Phi$ and $\chi$ denote the dilaton and axion fields, respectively, and $F_{(5)}$ is a five-form field strength coupled to the D3-branes. 
Under the ansatz $\chi=-e^{-\Phi}+\chi_0$, the dilaton and axion terms cancel each other in the above action and thus we obtain the solution with the metric and five-form field in the string frame as
\begin{align}\label{action22}
 ds^2_{10}&=e^{\Phi/2} \left\{\frac{r^2}{R^2}\left[- dt^2+d\vec{x}^2\right]+\frac{R^2}{r^2}dr^2+R^2d\Omega^2_5 \right\}, \nonumber  \\
 e^\Phi&=1+\frac{q}{r^4},~~~~ \chi=-e^{-\Phi}+\chi_0,
 \end{align}
where $R$ is the radius of the AdS space and $q$ represents the density of D-instantons, which corresponds to the vacuum expectation value of the gluon condensate in the gauge theory side. Here, the boundary is located at $r \to \infty$ and the temperature in the field theory side is zero. 
The metric of the $S^5$ can be written as $d\Omega^2_5=d\phi^2+\cos ^2\phi d \Omega^2_3+\sin ^2 \phi d\psi^2$.

Now, let us add the D7 probe brane to the bulk in the gravity side to introduce fundamental quarks to the field theory. 
The DBI action for the D7-brane is given by
\begin{align}\label{dbi}
 S_{D7}=-\tau_7 \int d^8 \xi \ e^{-\Phi}\sqrt{-\det \left(g_{ab}+2 \pi \alpha'F_{ab}\right)}+\tau_7\int d^8 \xi \frac{1}{8!}C_{(8)i_1\dots i_8},
 \end{align}
where $\tau_7=1/[g_s (2\pi)^7 \alpha'^4]$ is the tension of the D7-brane and $\xi_a$ denote the brane coordinates.
$g_{ab}=\partial_a x^M \partial_b x^N G_{MN}$ and $F_{ab}=\partial_a A_b- \partial_b A_a$ are, respectively, the induced metric and the field strength on the brane.
The second term in the right-hand side of the above equation which is the Chern-Simons term comes from the coupling of the axion field to the D7-brane. 
And, $C_{(8)}$ denotes the Hodge dual 8-form gauge potential of the axion field.

To describe the D7-brane embedding, it is more convenient to write the bulk metric as in the following form:
\begin{align}\label{bulkmetric}
 ds^2&=e^{\Phi/2} \left[\frac{r^2}{R^2}\left(- dt^2+d\vec{x}^2\right)+\frac{R^2}{r^2}\left(d\rho^2+\rho^2d\Omega^2_3 +dw^2+w^2d\psi^2\right)\right],
 \end{align}
where $r^2=\rho^2+w^2$.
We choose the D7-brane to be embedded along the directions ($t$,$\vec{x}$,$\rho$,$\Omega_3$).
Thus, the embedding of the D7-brane is specified by determining $w(\rho)$ and $\psi(\rho)$. 
By the use of symmetry, we can set $\psi=0$.
Then, the induced metric on the brane reads
\begin{align}\label{indmetric}
 ds_{D7}^2&=e^{\Phi/2} \left\{\frac{r^2}{R^2}\left(- dt^2+d\vec{x}^2\right)+\frac{R^2}{r^2}\left[\left(1+w'^2\right)d\rho^2+\rho^2d\Omega^2_3 \right]\right\}.
 \end{align}
In \cite{instanton2} the authors prove that by fixing the D7-brane along the $\psi$ direction, the Chern-Simons term in the DBI action becomes locally total derivative.
We have also presented their proof in the Appendix for the interested reader.
In the Appendix we have also discussed some of the effects of this term on the results found in this paper.
Consequently, it can be ignored completely and has no effect on the embedding of the D7-brane.
Therefore, the DBI action and the resulting equation of motion are obtained as
\begin{align}\label{DBIind}
S_{D7}=-\tau_7 V_3 \Omega_3\int dt d\rho \ e^{\Phi}\rho^3\sqrt{1+w'^2},
 \end{align}
and 
\begin{align}\label{DBIeq}
\frac{d}{d \rho}\left(\frac{e^{\Phi} \rho^2 w'}{\sqrt{1+w'^2}}\right)+\frac{ qw\rho^3e^{\Phi}\sqrt{1+w'^2}}{r^6}=0.
 \end{align}
This equation can only be solved numerically for an arbitrary value of $q$.
Solving this equation, we obtain the embedding function $w(\rho)$ which contains information about the physical quantities of the system.
According to the AdS/CFT dictionary, from the near-boundary expansion of the embedding function $w(\rho)$ one can obtain the quark mass $m_q$ and the chiral condensate $c(=\langle \bar{\Psi} \Psi\rangle)$, as follows:
\begin{align}\label{expansion}
w(\rho)=m_q+\frac{c}{\rho^2}+\dots.
 \end{align}

\begin{figure}[h]
\begin{center}
\includegraphics[width=6cm]{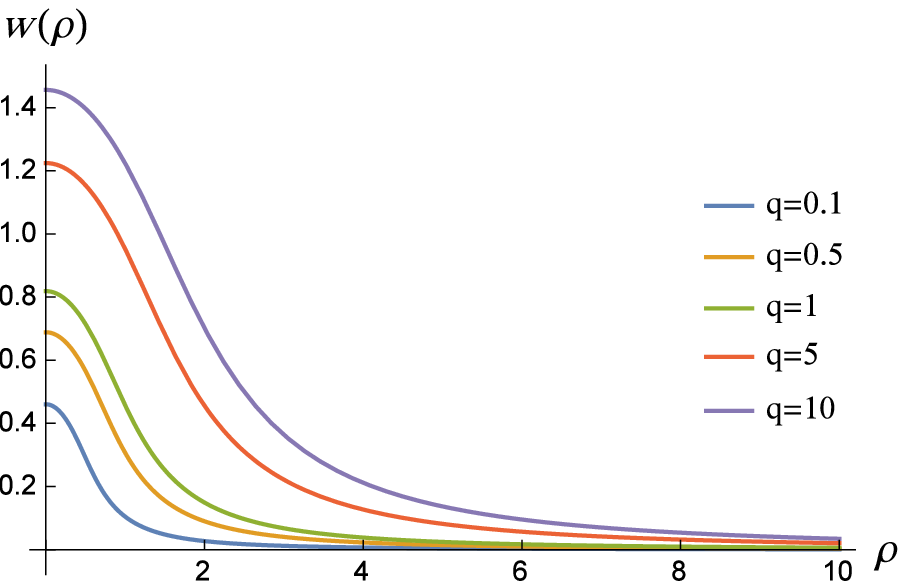}\hspace{1cm}
\includegraphics[width=6cm]{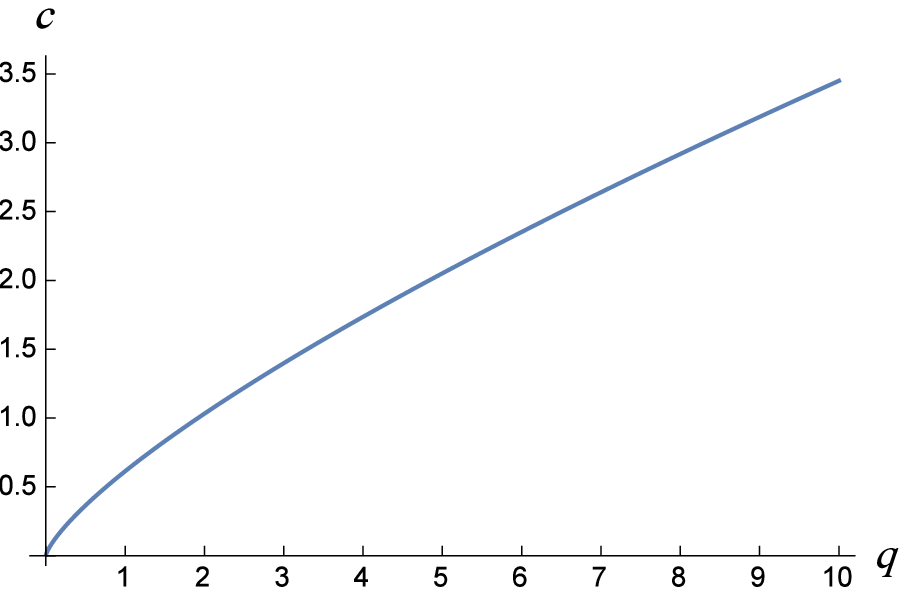}
\end{center}
\caption{\footnotesize 
Left graph: D7-brane embeddings for zero quark mass and different values of the gluon condensate.
Right graph: The quark condensate versus $q$ for zero quark mass.}
\label{emb-q}
\end{figure}

D7-brane embeddings for zero quark mass and different values of $q$ are depicted in the left graph of Fig.\,{\ref{emb-q}}.
As can be seen, the gluon condensate $q$ has the effect of repelling the D7-brane from the origin.
For $q=0$, the background is reduced to the $AdS_5\times S^5$ geometry and we obtain the flat embedding with $c=0$.
In the right graph of Fig.\,{\ref{emb-q}} we show the chiral condensate as a function of $q$.
From this graph, we observe that the gluon condensate increases the value of the chiral condensate and the chiral condensate is an almost linearly-increasing function of $q$, for large values of $q$.
The important result here is that the gluon condensate leads to the spontaneous chiral symmetry breaking, since at nonzero $q$ the chiral condensate is nonzero for the massless quarks\footnote{In \cite{linearP}, they keep the Chern-Simons term in their calculations, which leads to a completely different result. According to their calculations, the Chern-Simons term cancels the effect of the dilaton, resulting in flat D7-brane embedding and no chiral symmetry breaking.}.

\section{Response to an external electric field}\label{s3}
In this section, we investigate the effect of a static external electric field $E$ on the field theory with gluon condensate.
To describe such a system holographically, we need to introduce a gauge field in the form of  
\begin{align}\label{gauge}
A_x(t,\rho)=-E t+h(\rho)
\end{align}
on the D7-brane. 
Here, we have chosen the electric field to be in the $x$ direction of the field theory, without loss of generality.
Taking this gauge field into account, the DBI action is given by
\begin{align}\label{DBI-E}
S_{D7}=-{\cal N}\int dt d\rho \ e^{\Phi/2}\rho^3\sqrt{\left(1+w'^2\right)e^{\Phi}+(2 \pi \alpha')^2\left[ h'^2-\left(\frac{R}{r}\right)^4(1+w'^2)E^2\right]},
 \end{align}
where ${\cal N}=\tau_7 V_3 \Omega_3$.
From this action, the conserved charge associated with $h(\rho)$ reads
\begin{align}\label{charge}
j\equiv \frac{
 e^{\Phi/2}\rho^3(2 \pi \alpha')^2h'}{\sqrt{\left(1+w'^2\right)e^{\Phi}+(2 \pi \alpha')^2\left[ h'^2-\left(\frac{R}{r}\right)^4(1+w'^2)E^2\right]}},
\end{align}
where $j$ corresponds to the electric current in the $x$ direction of the field theory ($\langle J_x\rangle$).
As is obvious from the action (\ref{DBI-E}), $h'$ appears in the equation obtained for $w(\rho)$.
We can either find the equation of motion for $w(\rho)$ and then plug in $h'$ from Eq.\,(\ref{charge}) or Legendre transform the action (\ref{DBI-E}) to eliminate the gauge field and then derive the equation of motion for $w(\rho)$.
The Legendre transformed action has the following form:
\begin{align}\label{DBI-leg}
S_{D7}^L=-{\cal N}\int dt d\rho \left(\frac{R}{r}\right)^2\sqrt{1+w'^2} \sqrt{\left(\frac{q+r^4}{R^4}-(2 \pi \alpha')^2E^2\right)\left(\rho^6 e^{\Phi}-j^2 R^2\right)},
\end{align}
in which we have replaced $e^{\Phi}$ by $1+\frac{q}{r^4}$ in the first parenthesis.
After some calculations, the equation of motion for $w(\rho)$ is obtained as follows:
\begin{align}\label{eq}
\frac{d}{d\rho}\left(\frac{w'}{\sqrt{1+w'^2}}\sqrt{F(r)}\right)-\frac{w\sqrt{1+w'^2}}{2 r\sqrt{F(r)}}\frac{dF(r)}{dr}=0,
\end{align}
where $F(r)=\rho^6e^{2\Phi}\left(1-\frac{R^4(2 \pi \alpha')^2E^2}{r^4 e^{\Phi}}\right)\left(1-\frac{R^2j^2}{\rho^6e^{\Phi}}\right)$.
Now, using Eq.\,(\ref{DBI-leg}) and solving the equation of motion (\ref{eq}), we consider the behavior of the system under different values of the electric field and explore the possible phases.
\subsection{Phase transitions under the effect of the electric field}
One can simply observe that for given values of the electric field and instanton density, the action (\ref{DBI-leg}) vanishes at the radial position $r_s$ satisfying the equation
\begin{align}\label{rs}
 r_s^2=\rho_s^2+w(\rho_s)^2=\sqrt{R^4 (2 \pi \alpha')^2E^2-q}.
\end{align} 
As can be easily seen, in the absence of instantons, $q=0$, there always exists a singular radius for any nonzero value of the electric field.
However, for $q \neq 0$, there is a real $r_s$ only if
$E>E_s=\frac{1}{2 \pi \alpha'}\frac{\sqrt{q}}{R^2}$.
If $E<E_s$, there exists no singular radius in the bulk and the action (\ref{DBI-leg}) should be real along the whole radial direction. 
In this case the first parenthesis under the square root is positive at any radial position. 
Therefore, $j$ must be zero in order to keep the second parenthesis also positive at any $r$.
Due to this result, we call the case $E<E_s$, the insulator phase, in which the electric current is always zero.
However, when $E>E_s$, where there is a real $r_s$ at which the action vanishes, we need in general a nonzero value of $j$ to guarantee the reality of the action.
This phase is called the conductor phase.
It seems that there occurs an insulator-conductor phase transition at $E=E_s$.

$E_s=\frac{1}{2 \pi \alpha'}\frac{\sqrt{q}}{R^2}$ is the tension of the linear potential between the quark and antiquark in this theory \cite{linearP} and as we showed in \cite{dehghani}, it is the critical electric field below which the Schwinger effect cannot happen.
That is, no quark-antiquark pairs, even the massless ones, can be produced for $E<E_s$ and this is the reason why $j=0$ in this case.
As we argued there, there are two competing forces acting on quarks in such confined theories; an attractive confining force and a repulsive electric force growing with the value of the applied field.
For $E>E_s$ where the repulsion due to the electric force overcomes the confining force between the quarks, the production of the quark pairs of different masses becomes possible, depending on how much higher is the electric field than $E_s$.
Note that for deconfined theories (an example is our theory with $q=0$), $E_s=0$  and $j$ is nonzero for any nonzero value of the external electric field.
The quarks of given masses that their corresponding embeddings reach the singularity shell have nonzero $j$ and hence they are produced.
In the following, we want to completely explore this critical phenomenon and explain it using some graphs.
In all the results and graphs we set $R=1$ and $2\pi \alpha'=1/T_f=1$, where $T_f$ is the fundamental string tension.

As stated in the previous section, the instanton density has the effect of repelling the branes from the origin.
On the other hand, we know that 
the electric field has an opposite effect and in the presence of the electric field the D7-branes are attracted by the singularity caused by the electric field.
For a given electric field, by increasing the instanton density, the D7-branes are repelled from the origin more and more, until even the massless quarks cannot be produced.
\begin{figure}[h]
\begin{center}
\includegraphics[width=4.7cm]{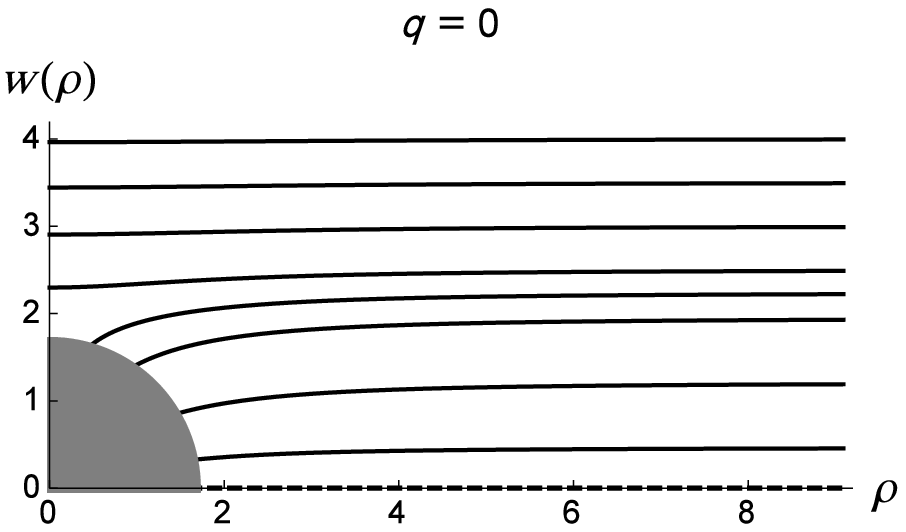}\hspace{0.1cm}
\includegraphics[width=4.7cm]{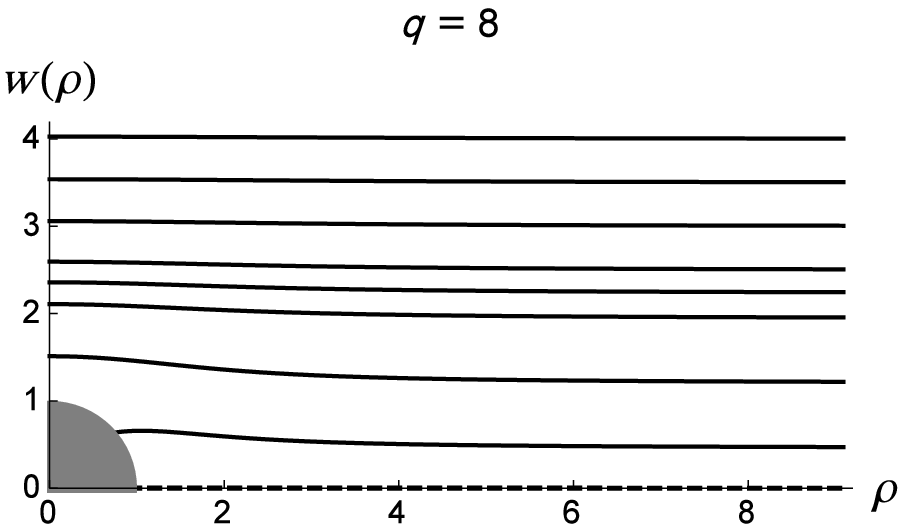}\hspace{0.1cm}
\includegraphics[width=4.7cm]{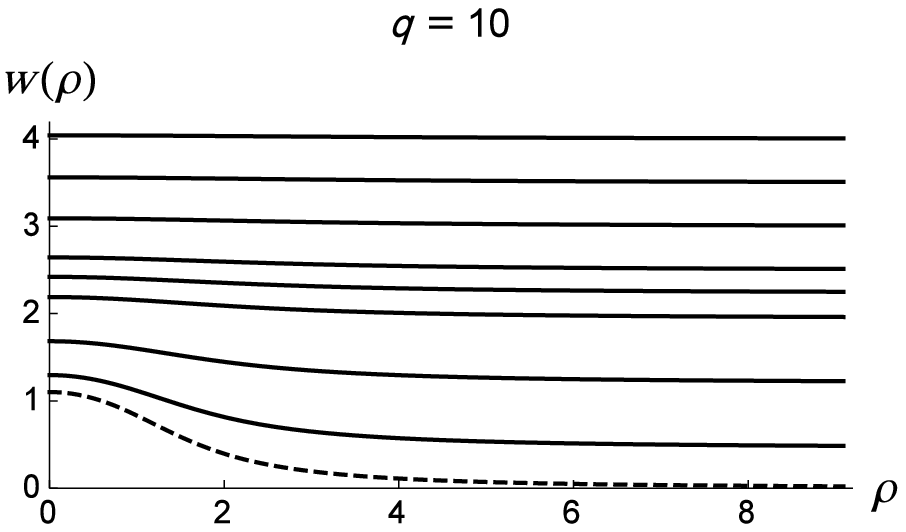}
\end{center}
\caption{\footnotesize 
D7-brane embeddings for $E=3$ and increasing values of $q$.
For large enough $q$, even the massless quarks, presented with the dashed lines, are not produced.
}
\label{higherq}
\end{figure} 
Figure \ref{higherq} shows the numerical results for the D7-brane embedding when the value of $q$ increases while the electric field $E$ is fixed.
According to these graphs, on the one hand $q$ repels the D7-brane and on the other hand it decreases the singularity radial position.
Both of these effects try to prevent the quark pairs from being produced. 
As can be seen from this figure, the embedding corresponding to massless quarks (represented by dashed line) is flat for small values of $q$.
However, when the instanton density increases so that the singular region disappears, as in the case in the right graph, it becomes nontrivial and the condensate is no longer zero for massless quarks.
This means that there is a phase transition; the increase of the instanton density, for any given $E$, leads to spontaneous chiral symmetry breaking.
At the same point, the phase of the system changes from the conductor to insulator. 
\begin{figure}[h]
\begin{center}
\includegraphics[width=6cm]{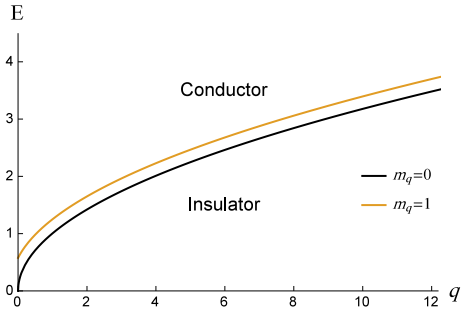}\hspace{1cm}
\end{center}
\caption{\footnotesize 
Phase diagram in the ($E$,$q$) plane. 
The black line shows the largest value of $q$, for a given $E$, for which the D7-brane embedding for massless quarks still reaches the singularity or exactly speaking there still exists a real singular region.
We have also depicted the result for $m_q=1$, for comparison.}
\label{phase}
\end{figure} 

The phase diagram is presented in Fig.\,\ref{phase}.
The black line is the critical line separating the insulator and conductor phases for massless quarks.
We also show the critical line for quarks of unit masses, for comparison.
In the insulator phase where there is no singular region, chiral symmetry is broken and in the conductor phase chiral symmetry is restored. 
The black line for the massless quarks is fitted to $E=\sqrt{q}$, which is the very critical electric field $E_s$, as expected.

\begin{figure}[h]
\begin{center}
\includegraphics[width=6cm]{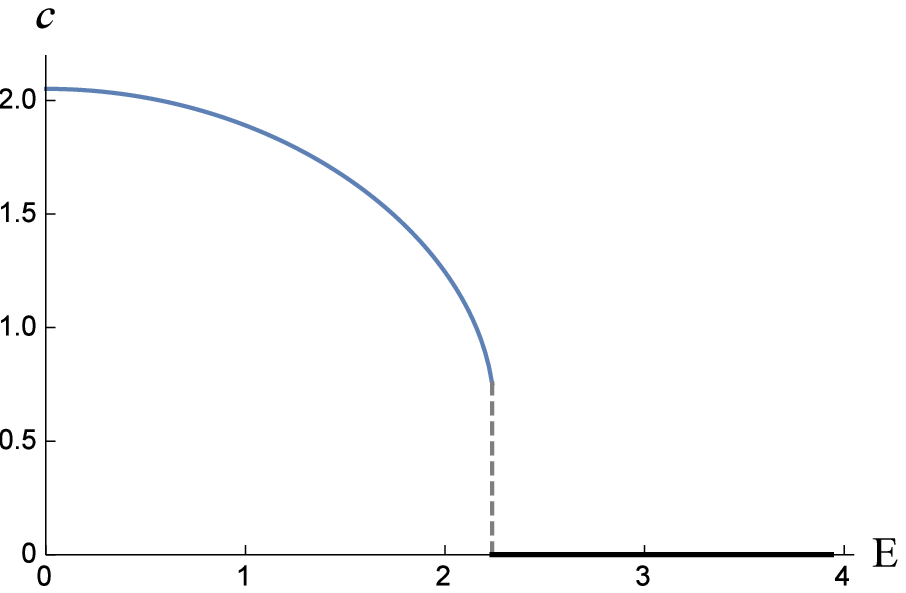}\hspace{1cm}
\includegraphics[width=6cm]{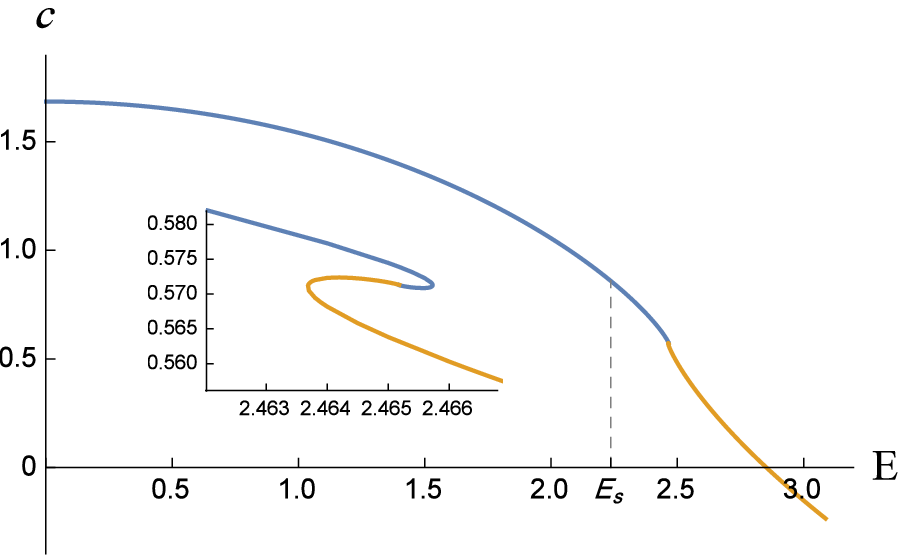}
\end{center}
\caption{\footnotesize 
Left and right graphs show the diagram of the quark condensate versus the electric field at $q=5$ for quarks of zero and unit masses, respectively.}
\label{phase2}
\end{figure} 
Now, let us elaborate further on the behavior of the quark condensate of the system under the changes of the electric field.
In the left graph of Fig.\,\ref{phase2} we draw the quark condensate versus the electric field for a fixed $q$ and massless quarks.
We can observe that by increasing the value of the electric field, the condensate for massless quarks decreases and right when the electric field approaches the critical electric field it suddenly vanishes, that is the chiral symmetry is restored in the conducting phase, as stated before. 

As one knows, for any field theory there is another critical electric field above which the Schwinger effect occurs without any restriction.
For the electric field less than this value the pairs can only be produced via a tunneling process.
To find 
this critical field in our calculations, in a theory with a fixed instanton density, $q=5$, we draw the diagram of the quark condensate for quarks of fixed mass, $m_q=1$, as a function of the external electric field. 
This diagram is shown in the right graph of Fig.\,\ref{phase2}.
From this graph we see that the condensate decreases when the electric field increases and no special behavior happens when passing through the critical field $E_s$. 
However, when we reach a larger electric field $E_c$, the behavior changes dramatically.  
As can be seen in the zoomed panel, a three-fold degeneracy is observed for a small range of fields and therefore a first order phase transition occurs at this point, as in other field theories studied before.
Note that the value of this critical field depends on the quark mass.

In the following, we study the behavior of the system in both the insulator and conductor phases, using more graphs.
\begin{itemize}[leftmargin=*]
\item 
{\bf Insulator phase ($E<E_s$):} As stated above, for $E<E_s$ there is no singular region and therefore only one kind of embedding (usually called Minkowski embedding) is possible for all the quark masses. 
Hence, the electric current is zero although the electric field is present.
The existence of such a critical electric field, below which no pairs even the ones of zero mass can be produced, is a common feature of confining theories.

\begin{figure}[h]
\begin{center}
\includegraphics[width=6cm]{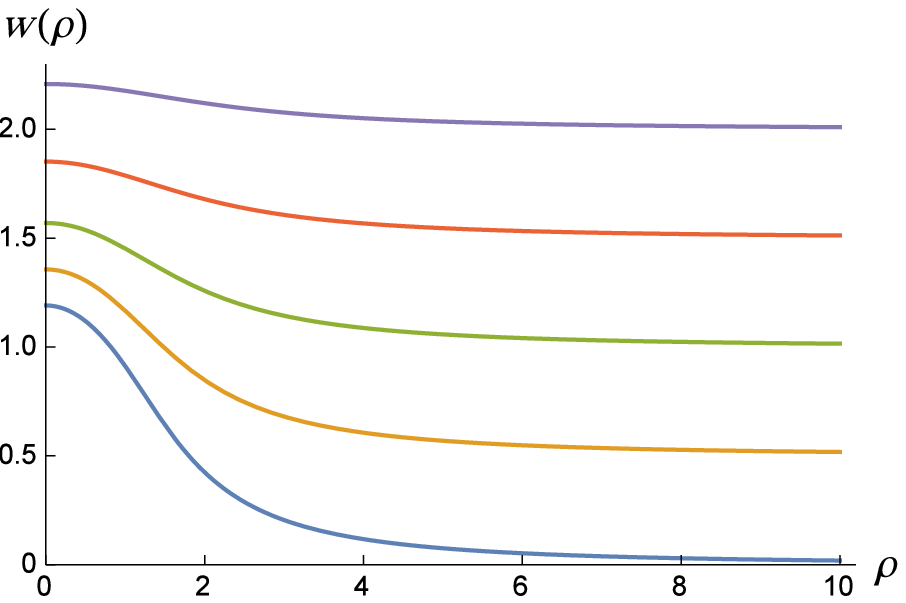}\hspace{1cm}
\includegraphics[width=6cm]{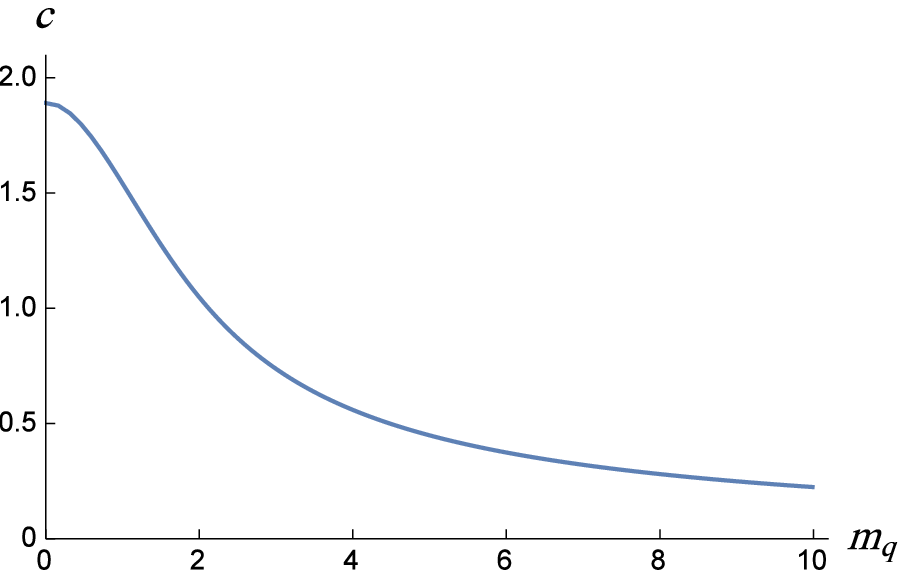}
\end{center}
\caption{\footnotesize 
Left graph: D7-brane embeddings for $q=5$, $E=1$, and $m_q=0,0.5,1,1.5,2$ from bottom to top.
Right graph: The quark condensate versus the quark mass.}
\label{emb-q-insol}
\end{figure} 

In the left graph of Fig.\,\ref{emb-q-insol} we depict the embeddings corresponding to quarks of different masses for given values of $q$ and $E$ (where $E<E_s$).
The right graph shows the quark condensate as a function of the quark mass for the same values of $q$ and $E$.
As can be seen, even the embedding for the zero-mass quark is nonflat and nontrivial, indicating a nonzero condensation, as is also obvious from the right graph.
This shows the spontaneous breaking of the chiral symmetry in this phase.
Note that the graph of the condensate versus the quark mass shows no jump or special behavior for $E<E_s$.
Thus, we find no phase transition in the insulator phase, in contrast to the result given in \cite{electric}.
They keep the Chern-Simons term in the DBI action of the D7-brane.
This leads to two results different from ours.
According to their calculations the flat embedding is preferred and chiral symmetry is not broken in this phase.
This result is not expected, since the zero mode of the fermions in the D-instanton
background requests chiral symmetry breaking. 
They also find a phase transition at some value of the quark mass, for which there is no reasonable and convincing physical interpretation.
As remarked before, the Chern-Simons term is shown \cite{instanton2} to become locally total derivative and consequently it cannot cancel the effect of the dilaton.

\item  
{\bf Conductor phase ($E>E_s$):}
For $E>E_s$ there exists a vanishing radius in the bulk and hence two kinds of embeddings are possible, depending on the corresponding quark mass.
The solutions for large enough quark masses do not cross the vanishing shell and the electric current is zero, that is although we are in the conductor phase, the repulsive force due to the electric field is not enough to overcome the color force between these heavy quarks.
However, lighter quarks can be produced. 
Solutions for these quarks reach the vanishing shell and the electric current for them is not zero\footnote{Strictly speaking, there are two types of these solutions for the D7-brane; those that satisfy $w(\rho=0)\neq 0$, which have a conical singularity and those with $w(\rho=0)=0$, which reach the origin without any singularity. These solutions are explained in many other cases, and first addressed in \cite{conical}.}. 

\begin{figure}[h]
\begin{center}
\includegraphics[width=6cm]{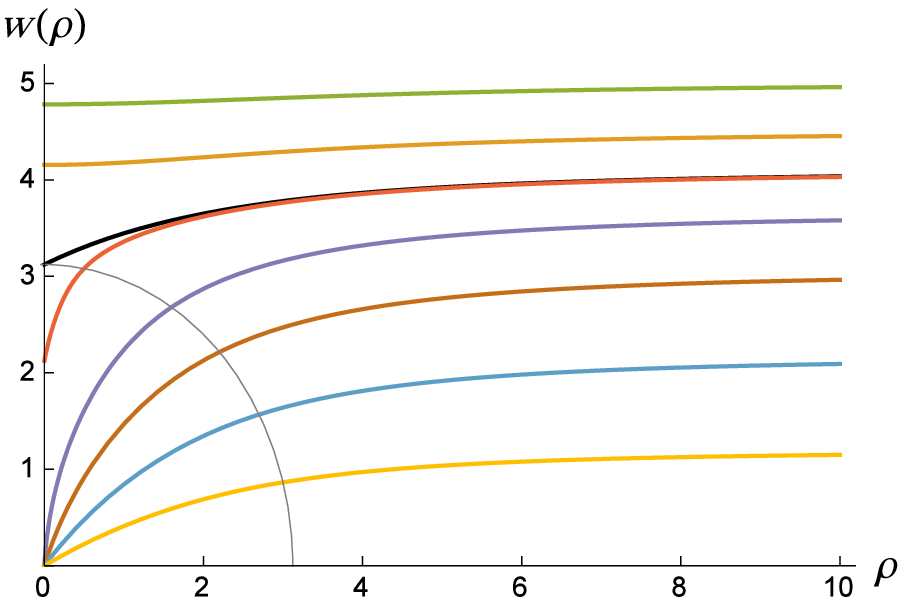}\hspace{1cm}
\includegraphics[width=6cm]{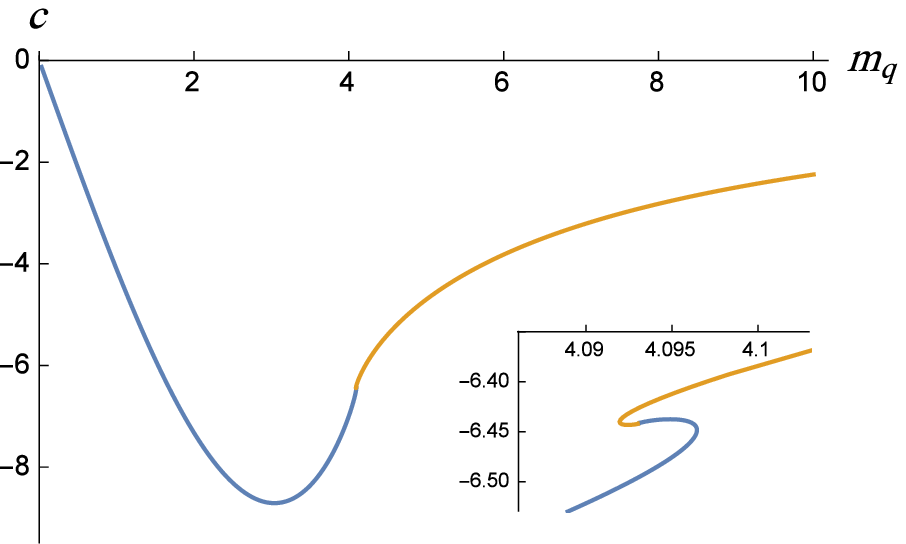}
\end{center}
\caption{\footnotesize 
Left graph: Different D7-brane embeddings for $q=5$, $E=10$, and $m_q=1.1943,2.1638,3.0447,3.654,4.0876,4.093,4.5,5$ from bottom to top. The gray curve shows the singular shell.
Right graph: The quark condensate versus the quark mass.}
\label{emb-q-cond}
\end{figure} 

A sample of the solutions for given values of $q$ and $E$ in the conductor phase is represented in Fig.\,\ref{emb-q-cond}.
The left graph shows some of the solutions with nonzero (zero) electric currents, which reach (do not reach) the singular shell.
In the right graph the quark condensate versus the quark mass is shown.
As can be seen, near the transition between the two types of the solutions, a multi-valuedness of the condensate is observed.
This result implies that there happens a first order phase transition, which is the aforementioned transition occurred at $E_c$ (see the right graph of Fig.\,\ref{phase2}). 
\end{itemize}

\section{Vacuum instability for the massless quarks}\label{imagi}
It is expected that an abrupt application of an electric field to a deconfined gauge theory causes its vacuum to become unstable.
In confined QCD-like theories, however, one expects a threshold electric field for the quarks even the massless ones be liberated and the Schwinger effect starts.
This is so, since a competing force is needed to overcome the confining force between quarks and antiquarks in a meson bound state.
It seems interesting to check this problem for the confined theory studied here.

The instability can be measured using the imaginary part of the effective Lagrangian.
This quantity shows the decay rate of the state of the gauge theory.
To evaluate it, we first find the effective DBI action in the presence of an electric field and serve it as the effective Lagrangian of the gauge theory by the use of AdS/CFT.
Suppose that we suddenly turn on an electric field on a solution of the system with $j=0$ and a specific embedding function $w(\rho)$, for a given $q$.
Notice that since we are concerned with the instability induced by the electric field, the solution over which we evaluate the decay rate is the solution to the embedding equation in the absence of the gauge field [Eq.\,(\ref{DBIeq})].

The effective Lagrangian in the presence of the electric field is
\begin{align}\label{eff-Lag}
{\cal L}_{\mathrm{eff}}(E)=-{\cal N}\int_0^{\infty} d\rho\ \frac{\rho^3 e^{\Phi/2}}{r^2}\sqrt{1+w'^2} \sqrt{r^4 e^{\Phi}-E^2},
\end{align} 
where $e^{\Phi}=1+\frac{q}{r^4}$.
It is evident that this integral becomes imaginary for $r<r_s$ where
\begin{align}\label{rs2}
 r_s^2=\rho_s^2+w(\rho_s)^2=\sqrt{E^2-q}.
\end{align} 
Hence, the interval of the integral can be broken into $r>r_s$ which gives the real part of the Lagrangian and $r<r_s$ which leads to the following form for the imaginary part:
\begin{align}\label{ImLag}
\Gamma_{\mathrm{eff}}(E)\equiv \mathrm{Im}{\cal L}_{\mathrm{eff}}(E)={\cal N}\int_0^{\rho_s} d\rho\ \frac{\rho^3 e^{\Phi/2}}{r^2}\sqrt{1+w'^2} \sqrt{r_s^4-r^4}.
\end{align} 

\begin{figure}[h]
\begin{center}
\includegraphics[width=7cm]{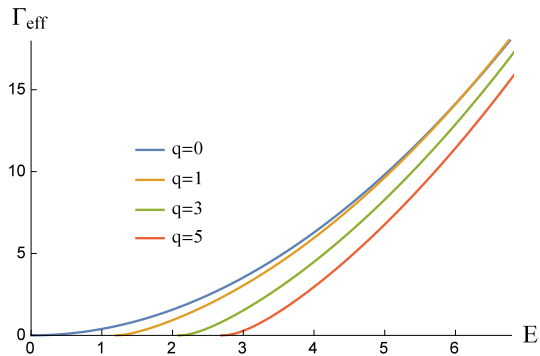}
\end{center}
\caption{\footnotesize 
The imaginary part of the effective Lagrangian of our theory with massless quarks versus the external electric field, for various values of the gluon condensate $q$.}
\label{decay}
\end{figure}

Figure \ref{decay} represents the imaginary part of the effective Lagrangian as a function of the electric field for several values of $q$. 
We have set ${\cal N}=1$.
It can be seen that $\Gamma_{\mathrm{eff}}$ becomes nonzero above the critical electric field $E_s$ explained in the previous section.
When $q=0$, however, the decay rate is nonzero only for nonzero electric fields, as expected for deconfined theories.
Moreover, one can observe that the decay rate is independent of the value of $q$ for large enough electric fields, although it extremely depends on $q$ when $E$ is small.
\section{Summary and conclusion}\label{conc}
The response of a confining field theory with nonzero gluon condensate to a background electric field has been considered using holography. 
The theory of our interest is a quasiconfining theory with confined quarks and deconfined gluons at zero temperature and becomes deconfined even for a small nonzero temperature. 
Adding a probe D7-brane to the gravity side, we have introduced the quarks to the theory and switched on a nontrivial $U(1)$ gauge field on the brane to describe the electric field in the dual field theory.

Changing the value of the electric field while keeping the other parameters fixed, we found two critical points in this theory.
One of them occurs when the electric field equals the string tension, i.e., the electric repulsion force between the quarks overcomes the confining color force between them. 
We observe that below this value, called $E_s$, no quark pairs of any masses are produced and therefore the electric current is zero.
Above this value the production of the quarks becomes possible.
Therefore, there is an insulator/conductor phase transition.
Moreover, another transition occurs at exactly the same point, which is the chiral symmetry restoration.
The theory at zero electric field has nonzero condensate even for zero-mass quarks, since instantons tend to repel the D7-brane from the origin.
Increasing the electric field from zero, decreases the condensate, since the electric field acts exactly in the opposite way.
In fact, the phase transition is the result of the interplay between these two effects.
When the electric field reaches its critical value, the chiral condensate for zero-mass quarks abruptly jumps to zero, showing a first order phase transition.

The other phase transition occurs at a higher value of the electric field $E_c$.
This one depends on the mass of the quarks as well as the instanton density of the system.
The phase transition at this point is revealed by studying the condensate for each quark mass as the electric field changes.
The behavior of this order parameter shows a first order phase transition at $E_c$.
Below this value for each quark mass, the production of these quarks is forbidden. 
However, lighter quarks can be produced, provided $E>E_s$.

In \cite{dehghani}, we found critical electric fields of the same system using a different approach based on the string worldsheet.
Calculating the total potential of quark-antiquark pairs of a given mass, we obtained two critical electric fields with an exactly same description as the ones obtained here using the DBI action.
We found the values of $E_c$ obtained from two approaches, however, to be different, while $E_s$s are exactly the same.
This result seems to be reasonable, since the string worldsheet and DBI action are different approximations of the string theory.
Some particular $\alpha'$ corrections are ignored in the DBI action and only a part of corrections are kept.
However, worldsheet calculations could capture different parts of the $\alpha'$ corrections.
They only may coincide in specific BPS cases or when some hidden symmetry is present, which are not the case here.
Therefore, the difference between the values of the critical electric field is indeed expected.
Since $E_s$ is determined by the string tension, it is found to be the same using two approaches.
However, $E_c$ as we found in our considerations, depends on the details of the theory and the quark mass, and thus it can be altered by the approximations that are used.

We have completed the study of the holographic Schwinger effect for our theory by calculating the full decay rate due to the onset of the instability caused by the presence of  the external electric field.
This has been done by calculating the imaginary part of the effective DBI action of the probe brane by taking the solution to the embedding equation obtained in the absence of the electric field, as the probe brane configuration.
As expected from the other parts of our study, the decay rate grows from zero only after the critical value of the electric field, $E_s$, is reached. 
Also, the decay rate is found to be independent of the gluon condensate as the electric field becomes large enough.

The study of quantum quenches in the context of quantum field theories is an important and challenging area of research.
Since there is lack of other theoretical techniques to deal with this problem, holographic techniques have been vastly used to investigate the time evolution of systems driven out of equilibrium under quantum quenches.
Recently there has been a growing interest in studying the quantum quenches in the confined phase (see e.g.\,\cite{quench-con1,quench-con2,quench-con3}), where they have found some evidence that the equilibrium (formation of a black hole in the gravity side) may not be the final fate of a system driven far from equilibrium by applying a quench.
In this regard, it would be interesting to study the evolution of a confining theory like the one in the present paper under electric field quenches, as an important class of quenches.
We would follow this path using appropriate numerical techniques \cite{ishii,equen,meson} in our future work.
\section*{Acknowledgement}
The authors would like to thank K. Hashimoto and R. Meyer for useful discussions. 
The work of Farid Charmchi
has been supported financially by Research Institute for Astronomy \&
Astrophysics ofMaragha (RIAAM) under research project No. 1/6025-62.
\appendix
\section{Chern-Simons action} \label{Calculation}

In this Appendix we show that if the position of the D7-brane is fixed along a fixed value of the $\psi$ direction, the Chern-Simons term in eq.\,(\ref{dbi}) becomes a locally total derivative term and thus does not play any role in the equation of motion.

When we embed the D7-brane in the bulk, the Hodge dual 8-form potential of the axion field can interact with D7-brane world volume.
To obtain the dual field, we introduce the vielbein
\begin{align}\label{A1}
&e^{\tilde{t}}=e^{\Phi/4}\frac{r}{R}dt,~~e^{\tilde{x}_i}=e^{\Phi/4}\frac{r}{R}dx_i,~~e^{\tilde{\rho}}=e^{\Phi/4}\frac{R}{r}d\rho,\nonumber\\
&e^{\tilde{\Omega}_3}=e^{\Phi/4}\frac{R}{r}\rho d\Omega_3,~~e^{\tilde{w}}=e^{\Phi/4}\frac{R}{r}dw,~~e^{\tilde{\psi}}=e^{\Phi/4}\frac{R}{r}wd\psi,
\end{align}
to write the metric (\ref{bulkmetric}) in terms of flat indices denoted by tilde.
Then, the field strength for the axion filed $\chi(r)$ and its Hodge dual are obtained, respectively, as follows:
\begin{align}\label{A2a}
F_{(1)}&=d\chi=\frac{\partial \chi}{\partial r}dr\nonumber\\
&=e^{-5\Phi/4}\frac{\partial \Phi}{\partial \rho}\frac{r}{R}e^{\tilde{\rho}}+e^{-5\Phi/4}\frac{\partial \Phi}{\partial w}\frac{r}{R}e^{\tilde{w}},\\
\label{A2b}F_{(9)}&=\frac{1}{9!}e^{-5\Phi/4}\frac{\partial \Phi}{\partial \rho}\frac{r}{R}\epsilon^{\tilde{\rho}}_{\tilde{t}\tilde{x}_i\tilde{\Omega}_3\tilde{w}\tilde{\psi}}
\ e^{\tilde{t}}\wedge e^{\tilde{x}_i}\wedge e^{\tilde{\Omega}_3}\wedge e^{\tilde{w}}\wedge e^{\tilde{\psi}}\nonumber\\
&+\frac{1}{9!}e^{-5\Phi/4}\frac{\partial \Phi}{\partial w}\frac{r}{R}\epsilon^{\tilde{w}}_{\tilde{t}\tilde{x}_i\tilde{\Omega}_3\tilde{\rho}\tilde{\psi}}
\ e^{\tilde{t}}\wedge e^{\tilde{x}_i}\wedge e^{\tilde{\rho}}\wedge e^{\tilde{\Omega}_3}\wedge e^{\tilde{\psi}}\nonumber\\
&=-\frac{4q \rho^4 w}{\left(\rho^2+w^2\right)^3} dt \wedge d \vec{x} \wedge d \Omega_3 \wedge dw \wedge d\psi \nonumber\\
&+ \frac{4q \rho^3 w^2}{\left(\rho^2+w^2\right)^3} dt \wedge d \vec{x} \wedge d \Omega_3 \wedge d\rho \wedge d\psi,
\end{align} 
where we have used $\epsilon_{\tilde{t}\tilde{x}_i\tilde{\rho}\tilde{\Omega}_3\tilde{w}\tilde{\psi}}=1$. In the final result we have used $e^{\Phi}=1+\frac{q}{r^4}$ and $r^2=\rho^2+w^2$.
Now, we can find the Hodge dual 8-form potential $C_{(8)}$ using $F_{(9)}=dC_{(8)}$.
Assuming
\begin{align}\label{A3}
C_{(8)}&=f(\rho,w,\psi) dt \wedge d \vec{x} \wedge d \Omega_3 \wedge dw+g(\rho,w,\psi)dt \wedge d \vec{x} \wedge d \Omega_3 \wedge d\rho
\end{align}
and then substituting $F_{(9)}$ from eq.\,(\ref{A2b}) and the relation of $dC_{(8)}$ into $F_{(9)}=dC_{(8)}$, we find the following conditions for $f(\rho,w,\psi)$ and $g(\rho,w,\psi)$:
\begin{align}\label{A4}
\frac{\partial f}{\partial \psi}=-\frac{4q \rho^4 w}{\left(\rho^2+w^2\right)^3},~~~
\frac{\partial g}{\partial \psi}=\frac{4q \rho^3 w^2}{\left(\rho^2+w^2\right)^3},~~~
\frac{\partial f}{\partial \rho}-\frac{\partial g}{\partial \psi}=0.
\end{align}
Integrating the two first relations we arrive at
\begin{align}\label{A5}
f(\rho,w,\psi)&=-\frac{4q \rho^4 w}{\left(\rho^2+w^2\right)^3}\left(\psi+\psi_1\right),\nonumber\\
g(\rho,w,\psi)&=\frac{4q \rho^3 w^2}{\left(\rho^2+w^2\right)^3}\left(\psi+\psi_2\right),
\end{align}
and using the last relation in eq.\,(\ref{A4}), we have $\psi_1=\psi_2=\psi_0$.
Therefore, if we fix the location of the D7-brane along the $\psi$ direction at $\psi=\psi_*$, the 8-form potential can be written as
\begin{align}\label{A6}
C_{(8)}=&=-\frac{4q \rho^4 w}{\left(\rho^2+w^2\right)^3}\left(\psi_*+\psi_0\right) dt \wedge d \vec{x} \wedge d \Omega_3 \wedge dw \nonumber\\
&+ \frac{4q \rho^3 w^2}{\left(\rho^2+w^2\right)^3} \left(\psi_*+\psi_0\right)dt \wedge d \vec{x} \wedge d \Omega_3 \wedge d\rho \nonumber\\
&=\left(\psi_*+\psi_0\right)d\left[\frac{q \rho^4}{\left(\rho^2+w^2\right)^2}dt \wedge d \vec{x} \wedge d \Omega_3\right].
\end{align}
This relation shows that $C_{(8)}$ is a pure gauge with zero field strength.
And, the Chern-Simons term is locally total derivative.
Therefore, it does not contribute to the equation of the DBI action and, as explained before, the Chern-Simons (CS) term cannot affect the embedding equation and should be removed completely.

Keeping this term and repeating the calculations, we can see that the presence of this term leads to some physically different results.
For instance, as stated before, this term cancels the Dilaton effect which in turn results in flat brane embedding and thus no chiral symmetry breaking.
However, such breaking of the chiral symmetry is expected since the zero mode of the fermions in the D-instanton background requests chiral symmetry breaking.

Another consequence of keeping this term can be observed when we apply an electric field and study the response of the theory.
The presence of the CS term in the calculations leads to a jump in the condensate \cite{electric} showing a new phase transition in the insulator phase ($E<E_s$) where we expect no special behavior and there is no convincing physical interpretation for such a transition.

Apart from these results, an important phase transition found in our calculation is absent if we do not remove the CS term.
In the presence of this term, condensation is zero for a massless quark regardless of the value of the electric field, meaning that the chiral symmetry is always present.
Therefore, in contrast to our result shown in the left graph of Fig.\,\ref{phase2}, no chiral symmetry transition would occur at any value of the external electric field.
This result contradicts the result of other confined theories such as Sakai-Sugimoto model \cite{sakai}.
The existence of such a chiral symmetry restoration found also in our calculations is physically explicable.
The attractive force responsible for finite $\langle \bar{\Psi} \Psi \rangle$ is eliminated by the electric repulsive force at $E=E_s$ and the condensation becomes abruptly zero.

We should mention here that all these differences in results are related to the fact that if we keep the CS term, we are working in inadequate coordinates that hide the energetically favorable solution with nonzero quark condensate for the zero-mass quark in the absence of the electric field.
This problem is removed when we choose the coordinates with $\psi=\psi_*$ for the D7-brane and all the following results regarding the presence of the electric field would be correct accordingly.

 \end{document}